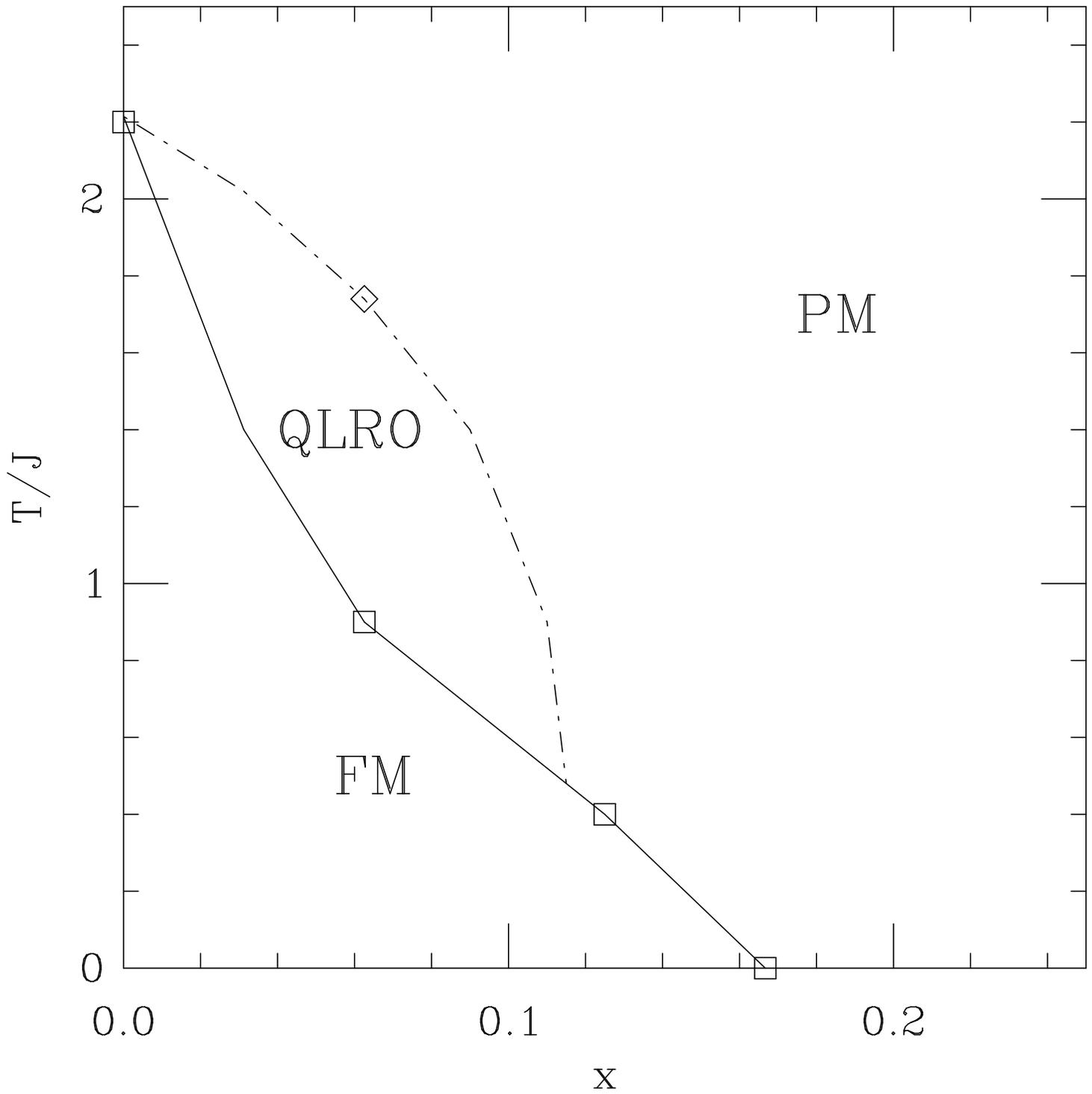

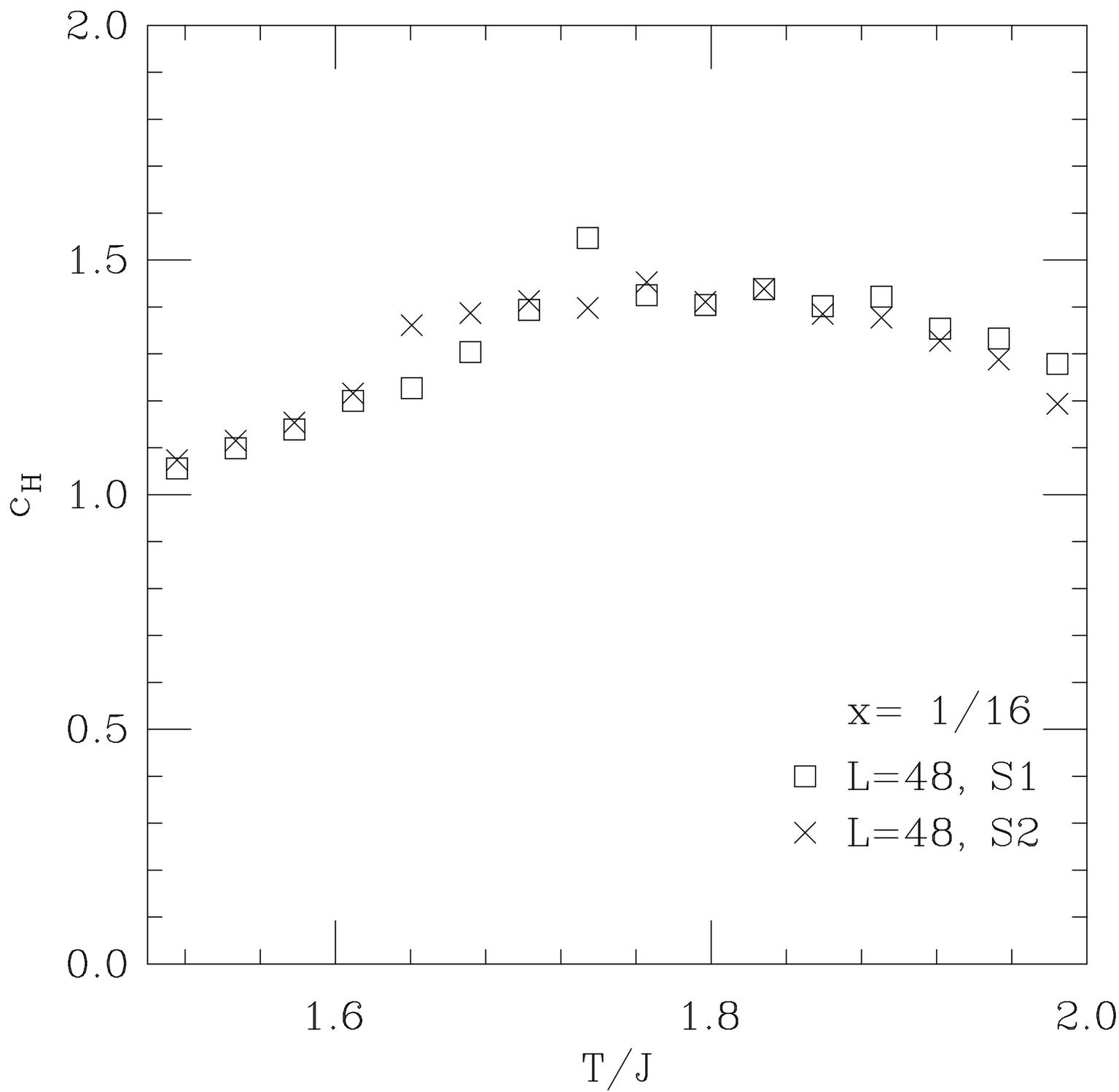

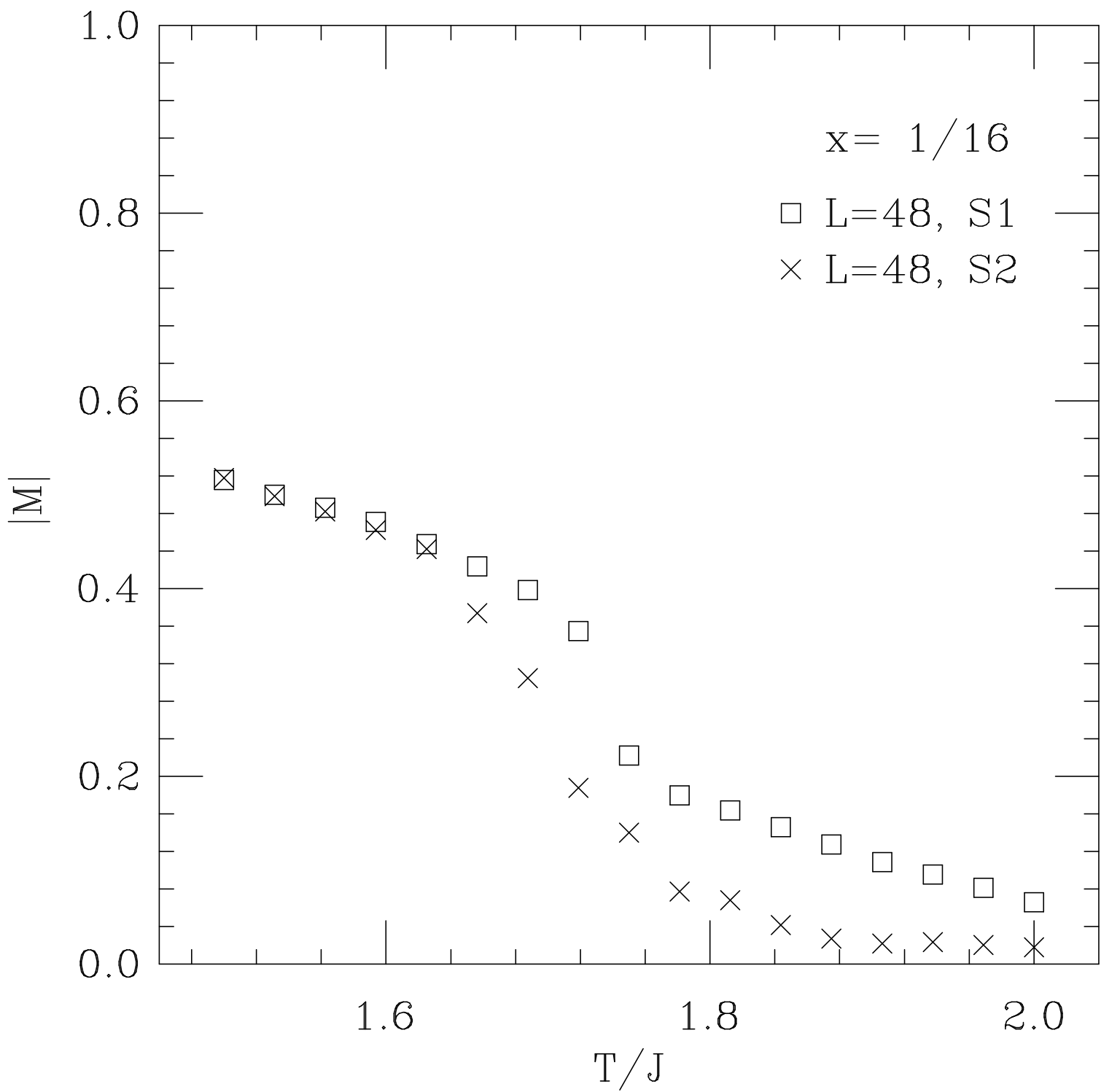

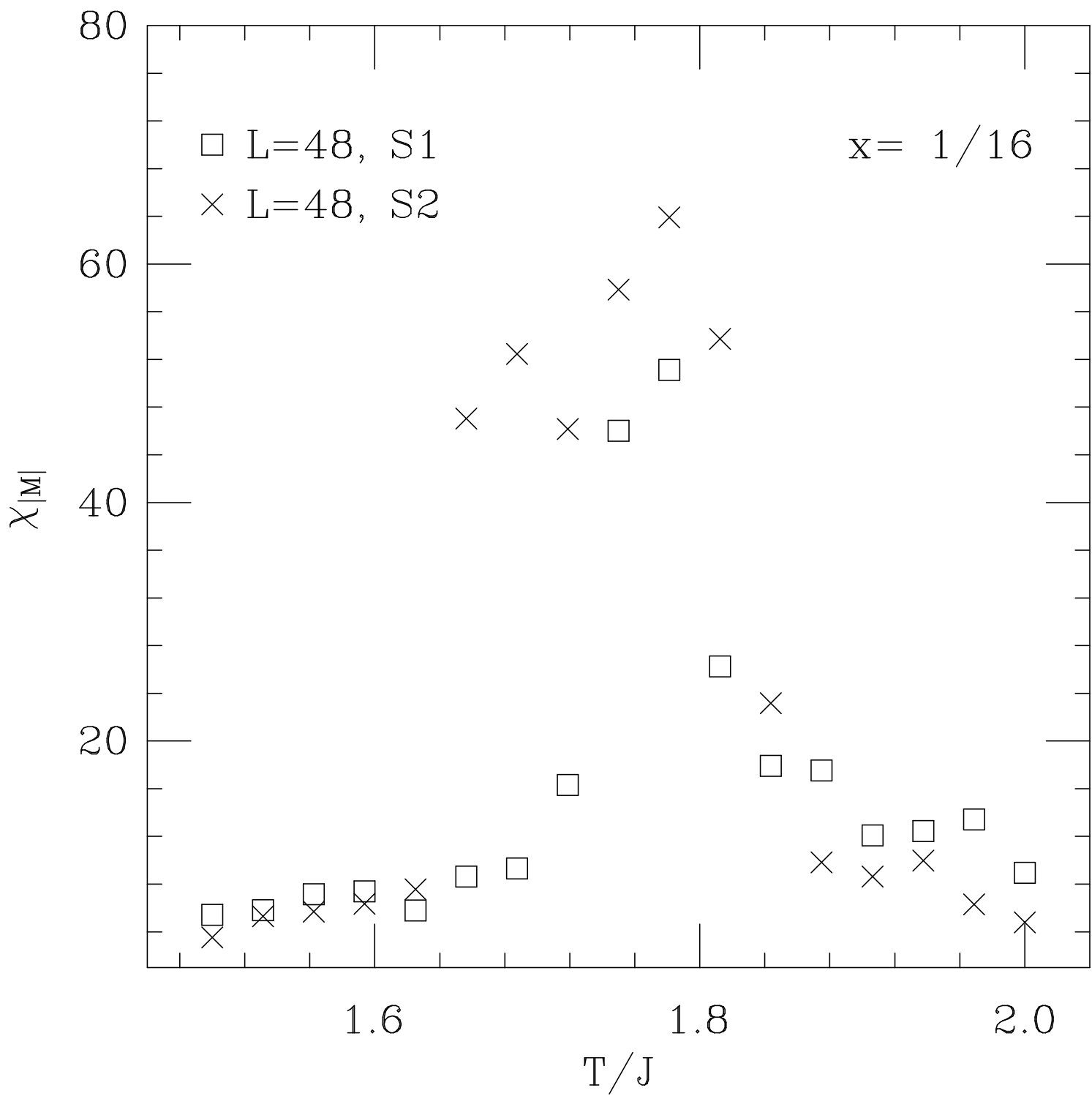

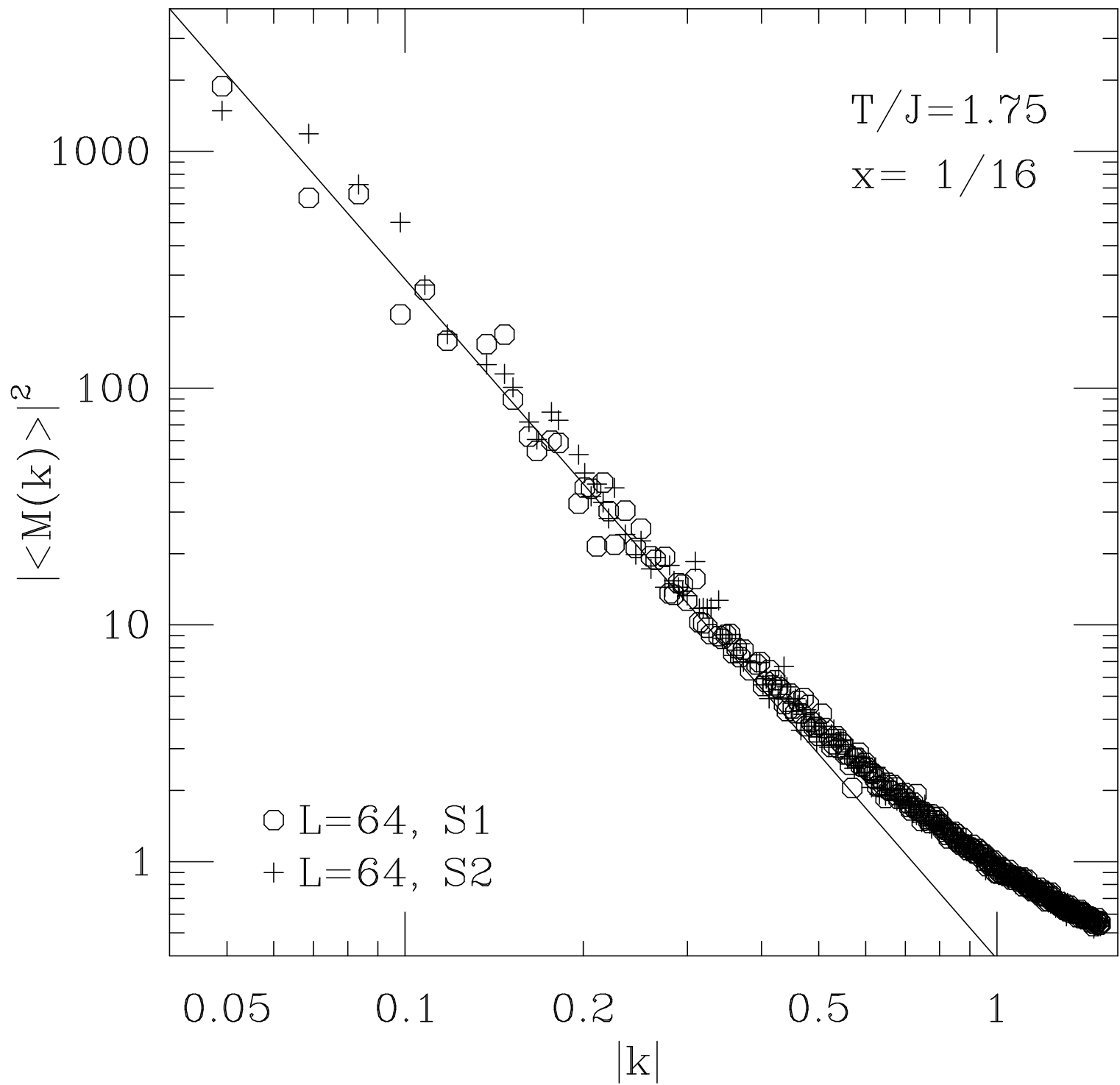

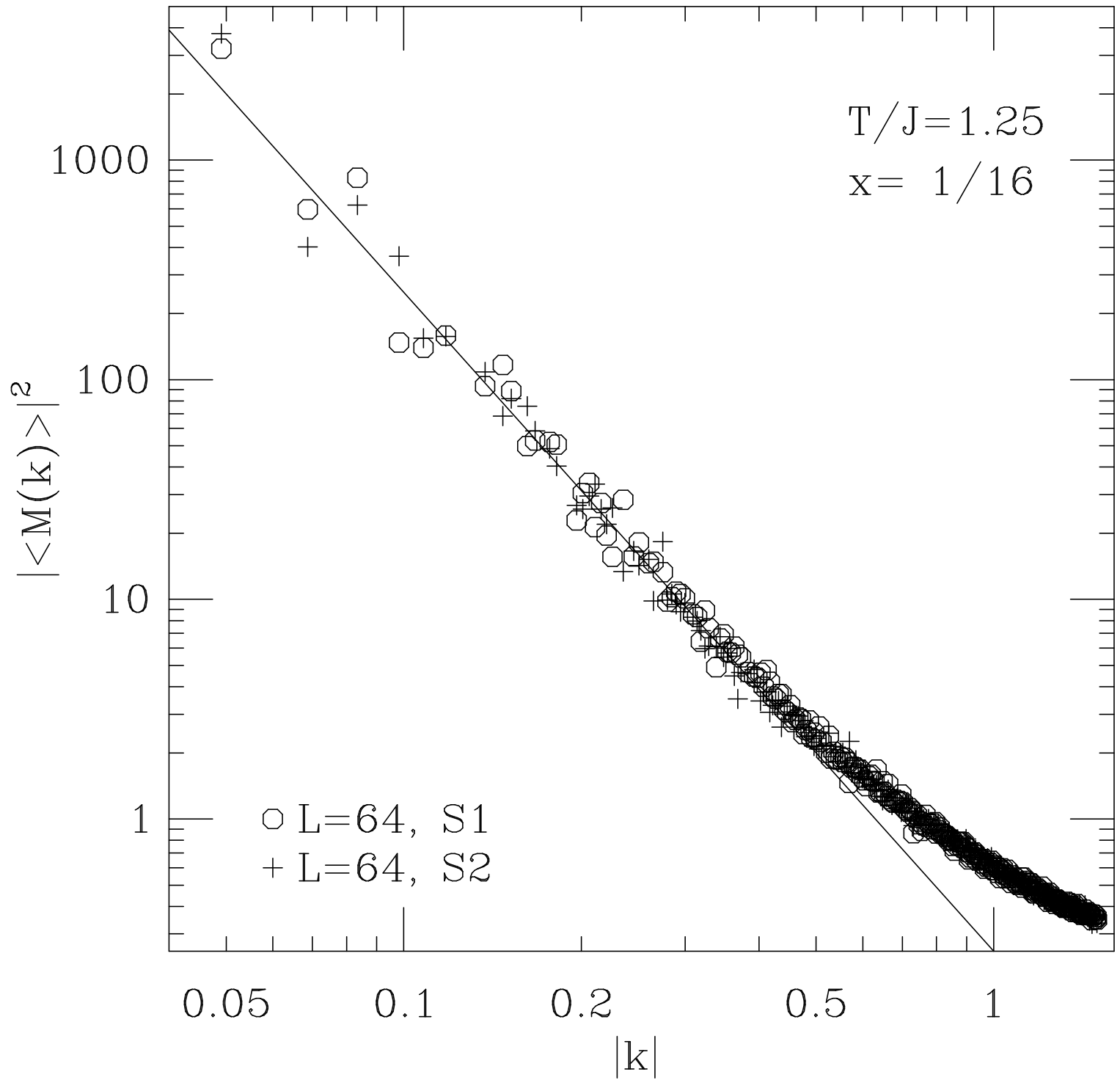

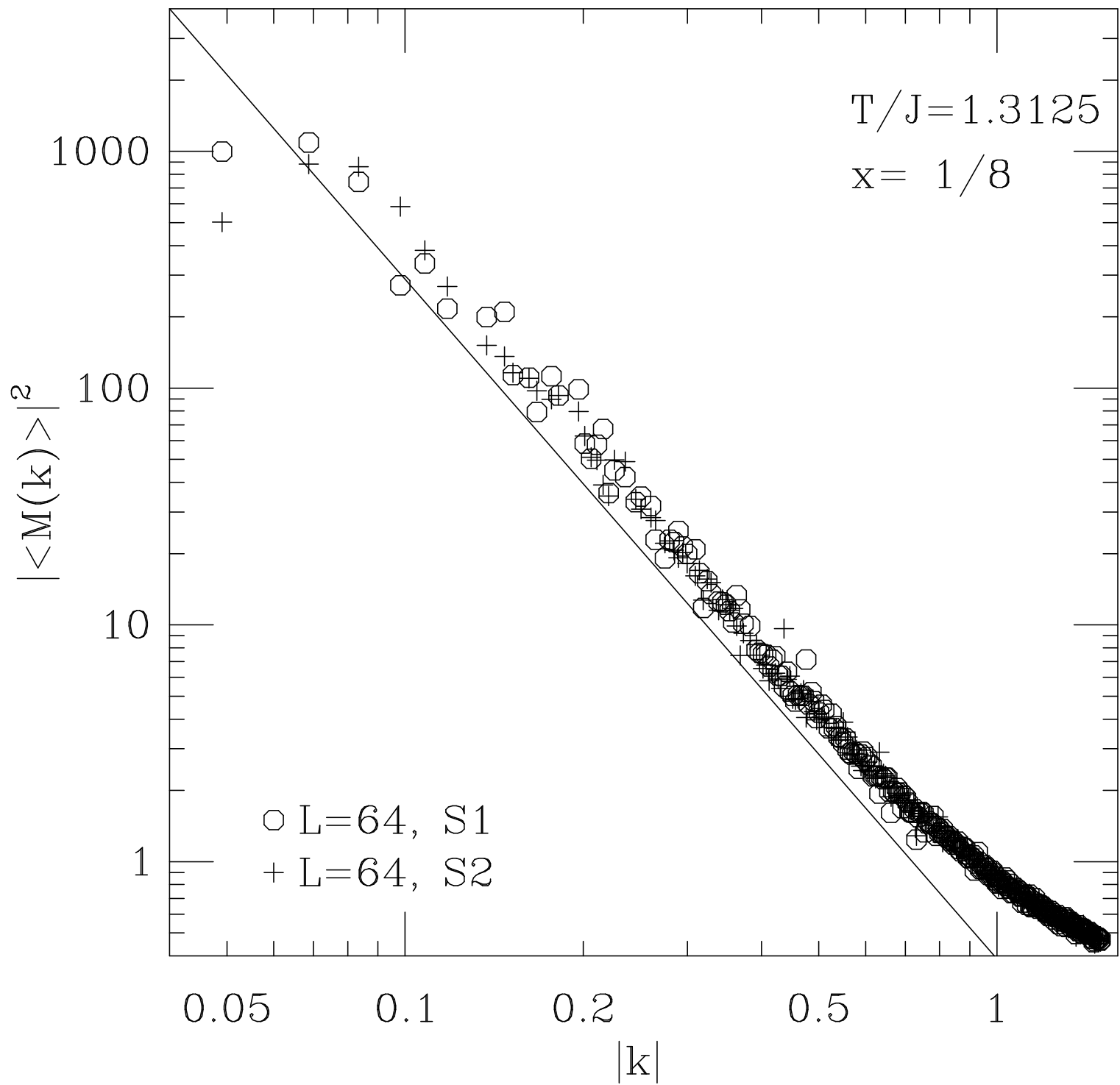

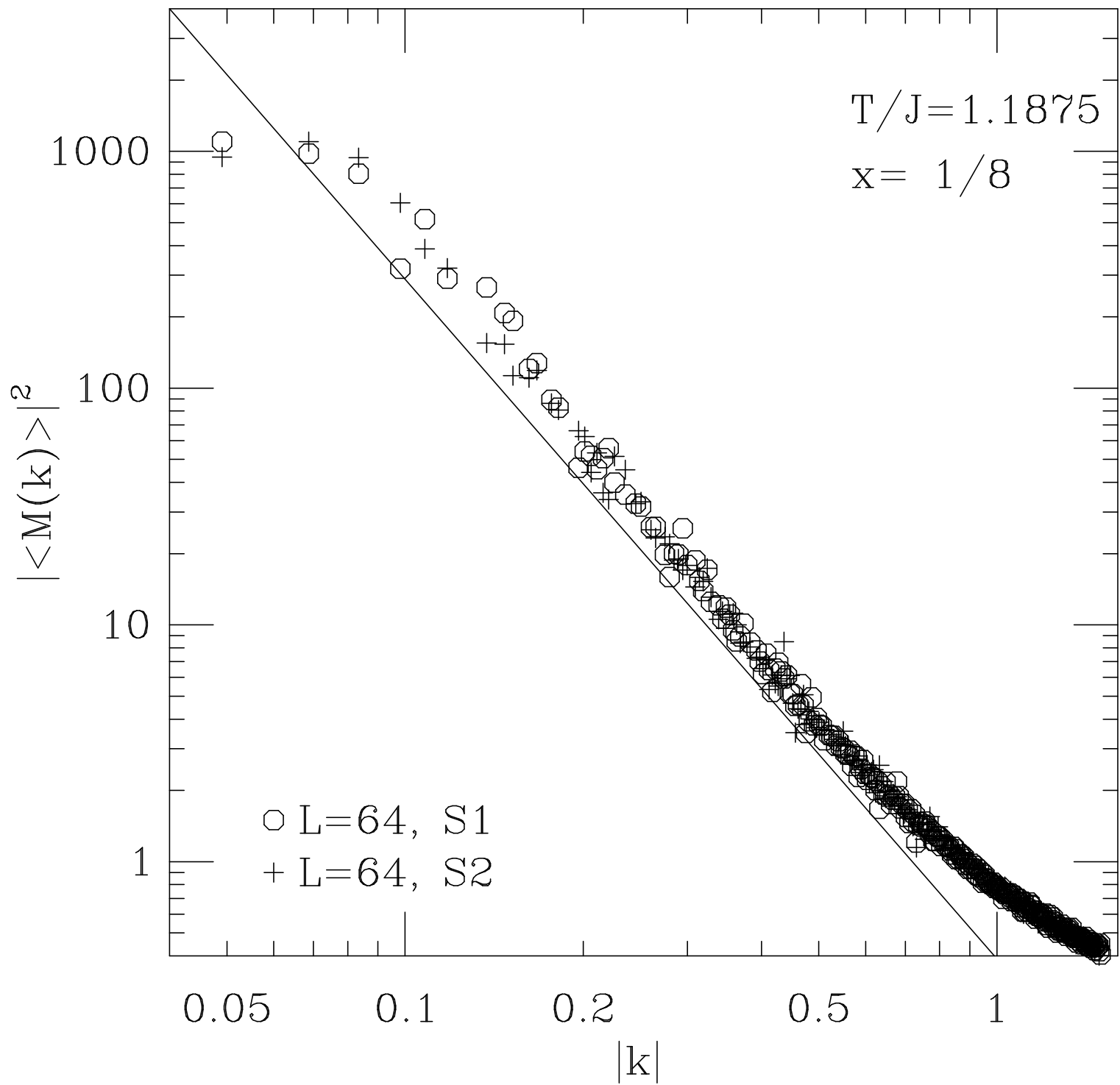

# POWER-LAW CORRELATED PHASE IN RANDOM-FIELD *XY* MODELS

# AND RANDOMLY PINNED CHARGE-DENSITY WAVES

Ronald Fisch
Dept. of Physics
Washington Univ.
St. Louis, MO 63130

ABSTRACT: Monte Carlo simulations have been used to study the $Z_6$ ferromagnet in a random field on simple cubic lattices, which is a simple model for randomly pinned charge-density waves. The random field is chosen to have infinite strength and random direction on a fraction $x$ of the sites of the lattice, and to be zero on the remaining sites. For $x= 1/16$ there are two phase transitions. At low temperature there is a ferromagnetic phase, which is stabilized by the six-fold nonrandom anisotropy. The intermediate temperature phase is characterized by a $|\mathbf{k}|^{-3}$ decay of two-spin correlations, but no true ferromagnetic order. At the transition between the power-law correlated phase and the paramagnetic phase the magnetic susceptibility diverges, and the two-spin correlations decay approximately as $|\mathbf{k}|^{-2.87}$. There is no evidence for a latent heat at either transition, but the magnetization seems to disappear discontinuously. For $x= 1/8$ the correlation length never exceeds 12, and the paramagnetic phase goes directly into the ferromagnetic phase; the two-spin correlation function is peaked at small $|\mathbf{k}|$, but the only divergence is the ferromagnetic delta function at $|\mathbf{k}| = 0$. The ferromagnetic phase terminates near $x= 1/6$.

## I. INTRODUCTION

It has been generally accepted for some time that a random field which couples linearly to the order parameter will always destroy the long-range phase coherence of a three-dimensional system which has a continuous symmetry. This result was first derived for the Abrikosov vortex lattice in type-II superconductors by Larkin,[1] and was later generalized to magnetic systems with random fields[2] and randomly pinned charge-density waves[3,4] (CDWs). This is believed despite the obvious fact that introducing the random field destroys the continuous symmetry, However, it has recently been questioned by Gingras and Huse,[5] who argue that these calculations have not taken proper account of the effects of vortex loops. By extrapolating Monte Carlo results from the strong random-field region, Gingras and Huse argue that at weak random fields there may be a vortex-free phase in which there is quasi-long-range order (QLRO), and the two-point correlations have a power-law decay as a function of distance. It is unclear what is intended by the phrase "vortex-free at large length scales". In three dimensions, vorticity is a vector, not a scalar. Even when there are no large vortex loops, the dipolar contributions of small vortex loops cannot be neglected. A three-dimensional *XY* model which is completely vortex-free is expected to be ferromagnetic.[6]

Based on renormalization group calculations, it was suggested by Mukamel and Grinstein[7] that it might be possible to induce a QLRO phase in a random-anisotropy system by adding a hexagonal crystalline anisotropy. This would seem to apply to the random-field case as well. In the work presented here we will adopt this suggestion. We will find that adding a six-fold anisotropy does allow us to find a QLRO phase, and we will study some of its properties. Since a CDW typically exists in the presence of some crystal potential, this result is directly applicable to cases of experimen-tal interest. We will not answer here the question of whether the QLRO phase exists even without the six-fold anisotropy. However, if it does exist in that case its properties should be essentially those that we find here.

We will assume that the amplitude fluctuations of the CDW are unimportant compared to the phase fluctuations.[3] This is reasonable because we expect short-range CDW order even in the high-temperature phase, so that only the phase coherence disappears at the critical temperature, $T_c$. Then, in a semi-classical lattice formulation, the theory may begin with the Hamiltonian

$$H = -J \sum_{<ij>} \cos(\theta_i - \theta_j - \mathbf{Q}\cdot(\mathbf{x}_i - \mathbf{x}_j)) - V \sum_i [\cos(n\theta_i) - 1] - G \sum_{i'} [\cos(\theta_{i'}) - 1]. \qquad (1)$$

The sum over $<ij>$ is a sum over all nearest neighbor pairs, and the sum over $i'$ is a sum over only defect sites, which are assumed to be randomly distributed and stationary. The $\theta_i$ variable represents the phase of the CDW at site $i$.

The $J$ term represents the stiffness of the CDW. The $V$ term, where $n$ is taken to be some integer, is a crude representation of the interaction of the CDW with a periodic crystal potential which wants the CDW to have a period commensurate with the lattice. In this work we will assume that $V$ is large, so that $\mathbf{Q}/n$ is equal to a Bravais lattice vector. As noted by Lee, Rice and Anderson,[8] when $n$ becomes large the difference between commensurate and incommensurate $\mathbf{Q}$ rapidly disappears, at least for the finite temperature behavior. The analogue for type-II superconductors of a commensurate CDW is a vortex lattice whose orientation in the plane perpendicular to the external magnetic field is locked in by the crystalline anisotropy.

The $G$ term gives the interaction of the CDW with the point defects. In this work we will study the strong pinning case of Fukuyama and Lee,[9] in which $G$ is taken to be large, and the fraction of defect sites, $x$, is taken to be small. In contrast to the one-dimensional case considered by Fukuyama and Lee, the strong pinning case becomes highly nontrivial in three dimensions when $x$ is small compared to $1 - p_c$, where $p_c$ is the site percolation concentration.[10]

## II. RANDOM FIELD MODEL

As is well known,[3] when we set $V = 0$ in Eq. (1) and make the gauge transformation $\theta_i \rightarrow \theta_i + \mathbf{Q} \cdot \mathbf{x}_i$ we obtain a random-field $XY$ model.[2] Explicitly, we have

$$H_{DRFXY} = -J \sum_{<ij>} \cos(\theta_i - \theta_j) - G \sum_{i'} [\cos(\theta_{i'} - \phi_{i'}) - 1], \qquad (2)$$

where $\phi_{i'} = -\mathbf{Q} \cdot \mathbf{x}_{i'}$. Since the defect sites are assumed to be immobile, the random fields do not change with time. As the random fields only occur on a fraction $x$ of the sites, Eq. (2) is more properly named the diluted random-field $XY$ model.

In the limit of large $V$ each $\theta_i$ is restricted to the values $2\pi l/n$, with $l = 1, 2, ... n$. As already noted, $\mathbf{Q}$ is commensurate in that limit, so the $\phi_{i'}$ are restricted to the same values. We then have a diluted random-field $Z_n$ model,[8,11]

$$H_{DRFZn} = -J \sum_{<ij>} \cos(\frac{2\pi}{n}(l_i - l_j)) - G \sum_{i'} [\cos(\frac{2\pi}{n}(l_{i'} - h_{i'})) - 1], \qquad (3)$$

where the random fields are now denoted by the $h_{i'}$ variables.

The $Z_2$ case, with random fields at every site ($x = 1$), is the much-studied random-field Ising model[12] (RFIM). An extensive Monte Carlo study of the $Z^3$ case has recently been reported by Eichhorn and Binder.[13] The $Z_4$ case reduces, as usual, to two independent $Z_2$ models, unless a four-spin term is added. In this work we study the $Z_6$ case. Eq. (3) becomes equivalent to Eq. (2) in the limit $n \rightarrow \infty$. Based on the argument of Lee, Rice and Anderson[8] and prior experience with the closely related random anisotropy problem,[14] we expect that in many respects the $Z_6$ case is already close to the large $n$ limit. This is known to be true in three dimensions both in the absence of any randomness and in the presence of random anisotropy.[14]

Because Eqs. (2) and (3) have no spatial symmetries, the existence of true phase transitions in these models is not entirely trivial. Even at high temperatures, the random fields will induce nonzero average values of the local site magnetizations. The existence of ferromagnetism in such systems requires that the exchange interactions be strong enough at low temperatures to prevent the system from breaking up into domains, as discussed by Imry and Ma.[2] The existence of a ferromagnetic phase transition requires, in addition, that there must be more than one low temperature minimum of the free energy in phase space (i.e. more than one Gibbs state), that the energy per particle of these minima be equal, and that averaging over these minima should restore the symmetry.

For Eq. (3), when $G/J$ is small it is natural to anticipate that the low-temperature behavior will be close to that of the $G = 0$ case. The Imry-Ma argument tells us to expect this for finite $n$ if the number of spatial dimensions is greater than two. This means that for a $Z_n$ model we should find $n$ ferromagnetic Gibbs states at low temperatures. Each of the $n$ ferromagnetic Gibbs states will have the same free energy per spin in the infinite volume limit, because the field treats all of the $n$ states equally, on the average. For the RFIM this was proven to be correct at $T = 0$ in three dimensions by Imbrie,[15] and the result was extended to finite temperature by Bricmont and Kupiainen.[16] The generalization of this result to any other finite value of $n$ is straightforward. However, the proof breaks down in the limit $n \to \infty$, because the domain wall energy (for rotation through a fixed angle) goes to zero as $1/n$.

For the three-dimensional RFIM,[12] there is a stable critical point, so that a (more or less) ordinary second order phase transition from the paramagnetic phase into the ferromagnetic phase is allowed. For $n > 4$ there is no known stable critical point in three dimensions, so either the transition is discontinuous, or else there should be an intermediate phase which possesses QLRO.[7] We will see that for the $Z_6$ model both cases occur, for different values of *x*. It is not obvious how many Gibbs states should exist in a QLRO phase. Since a QLRO phase requires an infinite correlation length, the two largest eigenvalues of the transfer matrix must become degenerate in the thermodynamic limit. Thus we expect at least two Gibbs states will exist in a QLRO phase. Without any spatial symmetry of the Hamiltonian, it is unlikely that the number of Gibbs states in a QLRO phase can become large for any value of *n*. The author believes that the number of such states will be either two or three. Unfortunately, obtaining a direct answer to this question by Monte Carlo simulation is probably impossible, because the lattice sizes required are too large.

### III. MONTE CARLO CALCULATION

As already mentioned, all of the Monte Carlo calculations reported here were done for the $n = 6$ case. In order to improve the efficiency of the computer program, the $G \rightarrow \infty$ limit of Eq. (3) was taken. The random field term then becomes a projection operator, which forces $l_{i'} = h_{i'}$ on all of the $i'$ sites, and the Hamiltonian reduces to

$$H_{Z6} = -J \sum_{<ij>} \cos(\frac{\pi}{3}(l_i - l_j)). \tag{4}$$

The projection operation is implemented by assigning a fraction *x* of the sites to be the $i'$ sites. The spins on these sites are given random directions, and then left fixed for the remainder of the calculation.

Eq. (4) has the useful property that the energy of every state is an integral multiple of 1/2. Thus it becomes possible to write a Monte Carlo program to study Eq. (4) which uses integer arithmetic to calculate energies. This gives substantial improvements in performance over working with the general form of Eq. (3), for both memory size and speed. It is also possible to use integer arithmetic if $G$ is chosen to be an integer.

The Monte Carlo program used two linear congruential pseudorandom number generators. In order to avoid unwanted correlations, the random number generator used to select which sites would be assigned the random fields was different from the one used to assign the initial $l_i$. A heat bath method was used for flipping the spins, which at each step reassigned the value of a spin to one of the six allowed states, weighted according to their Boltzmann factors and independent of the prior state of the spin.

$L \times L \times L$ simple cubic lattices with periodic boundary conditions were used throughout. The values of $L$ used ranged from 12 to 64, and typically two different random field configurations of a given $L$ were studied for a given $x$. This gave a rather crude estimate of the finite size dependence of the various thermodynamic properties. Unfortunately, high precision finite-size scaling is not a very effective tool for this problem, because the sample-to-sample variations for a given size are very large and not well behaved.[13,17] Because the errors in the transition temperatures we will calculate are dominated by the sample-to-sample fluctuations rather than the statistical errors of the data, we will not quote error estimates for $T_c$ in this work.

A remarkable property of the model studied here is that for interesting values of $x$ it is not plagued by the severe slow relaxations encountered for other field distributions which have been studied.[13,17] The author believes that this is because the dilute strong-pinning limit allows one to work at low average values of $\tanh(G/T)$ while remaining free of effects due to crossover to the pure (i.e. $G= 0$) system behavior.[17] For $x= 1/16$ the system is able to reach equilibrium in times comparable to those of the pure system at the same values of $L$ and the correlation length. This is in spite of the fact that, as we will see, we are clearly in a random-field dominated regime at $x=$ 1/16.

## IV. NUMERICAL RESULTS

A semi-quantitative picture of the phase diagram obtained from the Monte Carlo results is shown in Fig. 1. The ground state remains ferromagnetic for $x \leq 1/6$, with a magnetization along one of the six hexagonal directions. The magnetization seems to jump discontinuously to zero, although no quantitative estimates of the size of the jump were obtained. It is difficult to get precise estimates of the ferromagnetic transition temperature, $T_M$, because there are substantial differences in the stabilities of the six ferromagnetic minima. It is not clear to the author if one should estimate $T_M$ by looking at the most stable minimum or an average minimum. For $x= 1/16$ we find a critical temperature, $T_c= 1.74 J$, at which the magnetic susceptibility and the structure factor diverge. Between $T_c$ and $T_M$, which is less than 1.0 $J$ for this value of $x$, we find the QLRO phase. The $L$ dependence of the observed magnetization remains substantial down to $T= 1.0 J$.

The specific heat, $c_H$, of two $x= 1/16$, $L=48$ lattices is shown in Fig. 2. The data displayed were obtained by differentiating the calculated values of the energy with respect to $T$. The specific heat was also computed by calculating the fluctuations in the energy at fixed temperature, yielding similar but noisier results. Away from $T_c$ the samples were run for 10,240 Monte Carlo steps per spin (MCS) at each $T$, with sampling after each 10 MCS. Near $T_c$ they were run several times longer. The initial part of each data set was discarded for equilibration, as usual. We see that the data for the two samples agree fairly well, except in the region just below $T_c$. Despite the differences, the integrated specific heat of the two samples between $T= 1.5 J$ and $T= 2.0 J$ is essentially identical. There is a broad maximum in $c_H$ near $T_c$. It is clearly hopeless to try to estimate a value of the specific heat exponent $\alpha$ from these data, but it seems likely that $\alpha$ is less than $-1$.

The energy per site at $T = 1.75\ J$ is $-1.446\ J$. This is very low compared to the energy per site at the transition to QLRO in the random anisotropy model,[14] which is about $-1.014\ J$. This means that the nearest-neighbor spin correlations, which determine the energy of Eq. (4), are much larger at $T_c$ than for the random anisotropy model. It also means that the entropy, measured relative to the entropy at $T = \infty$, is much lower. The ground state energy is also much lower, about $-2.647\ J$, because $x$ is small.

The magnetization $<|M|>$ for the same two samples, over the same range of $T$, is shown in Fig. 3. The longitudinal magnetic susceptibility

$$\chi_{|M|} = \frac{1}{T}[<M^2> - <|M|>^2] \tag{5}$$

is shown in Fig. 4. The angle brackets indicate a thermal average. The existence of a non-zero value of $<|M|>$, even though $T > T_c$, is partly a finite size effect, and partly due to the random field. Although the behavior of the two samples is qualitatively similar, the quantitative differences are obvious.

Although we have seen that the sample-to-sample fluctuations prevent us from using finite-size scaling to obtain quantitative estimates of critical exponents, we can get valuable information by looking at the structure factor of samples of size $L=64$. The structure factor is the spatial Fourier transform of $<M^2>$, and it can be measured by X-ray or neutron scattering experiments. Near a critical point the long-wavelength behavior of the structure factor of a random-field model is expected to have the form

$$|<M(\mathbf{k})>|^2 \approx (1/\xi^2 + |\mathbf{k}|^2)^{-(4-\bar{\eta})/2}. \tag{6}$$

In three dimensions $\bar{\eta} \geq 1$. The correlation length $\xi$ becomes infinite at $T_c$. To estimate $\bar{\eta}$, we measure the slope of the structure factor on a log-log plot. This is shown, averaged over angles, for two $L=64$ lattices with $x= 1/16$ at $T= 1.75$, in Fig. 5(a). The slope of the best fit to the data is $-2.87 \pm 0.05$, so we find

$$\bar{\eta} = 1.13 \pm 0.05 . \qquad (7)$$

This value is indistinguishable from the value of $\bar{\eta}$ found by Eichhorn and Binder[13] for the random-field $Z_3$ model, but it is somewhat greater than the value found for the RFIM by Rieger and Young.[17] The author believes this indicates that the number of Gibbs states in the QLRO phase is probably three.

The correlation length remains infinite everywhere in the QLRO phase (essentially by definition). Repeating the above procedure for the same two lattices at $T= 1.25$, we find the results shown in Fig. 5(b). We see that the structure factor again shows a power-law behavior at small $|\mathbf{k}|$, but that the slope has now assumed the maximum allowed value of 3. Data for $T= 1.00$ (not shown) yields the same result. Thus we find that the value of $\bar{\eta}$ inside the QLRO phase, which we call $\bar{\eta}_0$, is

$$\bar{\eta}_0 = 1.00 . \qquad (8)$$

A value of $\bar{\eta}_0= 1$ precludes the possibility that the system is ferromagnetic in this range of $T$.

With $x= 1/8$, the results for small lattices lead one to expect that there is a critical point near $T= 1.25\ J$. The structure factor for the $L=64$ lattices tells a different story in this case, however. In Fig. 6 we show the results for two $x= 1/8$, $L=64$ lattices at $T= 1.3125\ J$ and at $T= 1.1875\ J$. We find that $\bar{\eta}$ is consistent with the value of 1.13 found in Fig. 5(a). However, for $x= 1/8$ we do not find a divergence of $\xi$; instead it seems to saturate at a value of about 12 lattice units. Therefore we conclude that when $x= 1/8$ the random field has become too strong to allow the QLRO phase to exist. The $T= 1.3125\ J$ data were obtained using a hot start condition and the $T= 1.1875\ J$ data used a cold start. Similar results were found at $T/J= 1.125$ and $1.0625$.

The author does not believe that the finite value of $\xi$ is due to a lack of equilibration. It should be noted that the value of $|<M(0)^2>|$, which is not shown on a log-log plot, is larger at the lower temperature.

## V. DISCUSSION

It might be objected that there is no real divergence of $\xi$ for $x=1/16$, and that the correlation length has merely become comparable to the size of our samples. The author believes this to be unlikely. An Imry-Ma argument predicts that in three dimensions if $x$ is reduced by a factor of two, then $\xi$ should increase by the same factor. A correlation length of 24 is not consistent with the data shown in Fig. 5(a). Further, if $\xi$ was finite for $x=1/16$, then we would expect the value of $\bar{\eta}$ to remain unchanged as we decrease $T$, as happens for $x=1/8$. This is not in agreement with the data shown in Fig. 5(b).

What will happen to the phase diagram in three dimensions as we increase $n$? While we a not in a position to give a definitive answer to that question, the author's opinion is the following. As we increase $n$ the region of stability of the ferromagnetic phase will shrink, both in $T$ and in $x$. In the limit $n\rightarrow\infty$, ferromagnetism is probably restricted to $x=0$. Based on prior experience with the random anisotropy model,[14] the author believes that the paramagnet-QLRO phase boundary will be almost unchanged, except that for large $n$ it will extend to $T=0$, instead of intersecting the ferromagnetic phase. It would certainly be desirable to check this by doing Monte Carlo calculations for larger values of $n$.

Given the results presented here, the similar results for the three-dimensional random anisotropy model, and the fact that high-temperature series expansions[19] for random anisotropy indicate that, in that case, the ferromagnetic phase remains stable in four dimensions, we are led to the possibility that the lower critical dimension for random-field models with continuous symmetry is actually three rather than four. Even if the QLRO phase does exist for Eq. (2) in three dimensions, as we have argued is likely, we have no direct evidence for a stable ferromagnetic phase in the four-dimensional random-field *XY* model. Still, we should ask why the Imry-Ma argument is so widely believed, since its lack of rigor is transparent. In the author's opinion, the primary support for the Imry-Ma argument for continuous spins is the idea of dimensional reduction.[20] This shows that the $\varepsilon$-expansion for an *m*-component spin model in the presence of the random field around six dimensions is identical to the expansion around four dimensions in the absence of the random field, and thus predicts that the lower critical dimension for $m \geq 2$ is four. However, the author can see no reason why dimensional reduction should be more accurate for $m \geq 2$ than it is for $m = 1$, where it is known[15,16] to be incorrect. It seems to the author entirely reasonable that dimensional reduction should make an error of equal size in both cases, which would be the case if the lower critical dimension was three when $m \geq 2$.

## VI. CONCLUSION

In this work we have used Monte Carlo simulations to study the $Z_6$ version of the diluted random-field ferromagnet in three dimensions. This is also a model of randomly pinned charge-density waves, and related to the randomly pinned Abrikosov lattice. We have found that, as suggested by the calculation of Mukamel and Grinstein, there are two ordered phases. In addition to the anisotropy-stabilized ferromagnet, for a sufficiently dilute concentration of pinning sites we find an intermediate phase displaying a power-law decay of two-spin correlations. We have obtained the critical exponent $\bar{\eta}$, which characterizes this power law on the critical line, and also the exponent $\bar{\eta}_0$ which is observed within the QLRO phase. Calculations which probe more details of the properties of this QLRO phase would clearly be desirable.

**Acknowledgment**

The author thanks David Huse and Michel Gingras for discussions of their work on random *XY* models, which encouraged the author to undertake the calculations described here.

FIGURE CAPTIONS

Figure 1. Phase diagram of the dilute random-field $Z_6$ model with $G=\infty$, showing the paramagnetic (PM), ferromagnetic (FM), and quasi-long-range order (QLRO) phases. The plotting symbols show estimates obtained from the Monte Carlo data, and the lines are guides to the eye.

Figure 2. Specific heat near $T_c$ of the dilute random-field $Z_6$ model with $x= 1/16$ on two $L=48$ simple cubic lattices.

Figure 3. Magnetization near $T_c$ of the dilute random-field $Z_6$ model with $x= 1/16$ on two $L=48$ simple cubic lattices.

Figure 4. Longitudinal magnetic susceptibility of the dilute random-field $Z_6$ model with $x= 1/16$ on two $L=48$ simple cubic lattices. $\chi_{|M|}$ is defined in Eq. (5).

Figure 5. Angle-averaged two-spin correlation function for the dilute random-field $Z_6$ model with $x= 1/16$ on two $L=64$ simple cubic lattices, log-log plot. Each data set shows averaged data from 2 states sampled at 10,240 MCS intervals. (a) $T= 1.75$, the line has a slope of $-2.87$; (b) $T= 1.25$, the line has a slope of $-3.00$. Note that the vertical scales differ in (a) and (b).

Figure 6. Angle-averaged two-spin correlation function for the dilute random-field $Z_6$ model with $x= 1/8$ on two $L=64$ simple cubic lattices, log-log plot. Each data set shows averaged data from 2 states sampled at 10,240 MCS intervals. (a) $T= 1.3125$; (b) $T= 1.1875$. The scales and lines are identical to the ones shown in Fig. 5(a).